\input ./style/arxiv-general.cfg
\documentclass[aos,MSNbibl,nameyear,rotating,dvips]{arximspdf}
\makeatletter
   \@ifpackageloaded{graphicx}{}{\usepackage{graphicx}}
\makeatother

\doi{10.1214/15-AOS1325}
\volume{43}
\issue{4}
\pubyear{2015}
\firstpage{1801}
\lastpage{1830}
\docsubty{FLA}

\makeatletter
\newcommand{\rrVert}{\Vert}
\newcommand{\llVert}{\Vert}
\newcommand{\Ig}{\mathcal{I}_g}

\newtheorem{theorem}{Theorem}
\newtheorem{lemma}{Lemma}
\newtheorem{corollary}{Corollary}
\newtheorem{assumption}{Assumption}
\makeatother

\begin{document}
\begin{frontmatter}

\title{Maximin effects in inhomogeneous large-scale data}
\runtitle{Maximin effects}

\begin{aug}
\author[A]{\fnms{Nicolai}~\snm{Meinshausen}\corref{}\ead[label=e1]{meinshausen@stat.math.ethz.ch}}
\and
\author[A]{\fnms{Peter}~\snm{B\"uhlmann}\ead[label=e2]{buhlmann@stat.math.ethz.ch}}
\runauthor{N. Meinshausen and P. B\"uhlmann}
\affiliation{ETH Z\"urich}
\address[A]{Seminar f\"ur Statistik\\
ETH Z\"urich\\
R\"amistrasse 101\\
CH-8092 Z\"urich\\
Switzerland\\
\printead{e1}\\
\phantom{E-mail: }\printead*{e2}}
\end{aug}

%
\received{\smonth{6} \syear{2014}}
%
\revised{\smonth{11} \syear{2014}}

%
\begin{abstract}
Large-scale data are often characterized by some degree of
inhomogeneity as data are either recorded in different time regimes or
taken from multiple sources. We look at regression models and the
effect of randomly changing coefficients, where the change is either
smoothly in time or some other dimension or even without any such
structure. Fitting varying-coefficient models or mixture models can be
appropriate solutions but are computationally very demanding and often
return more information than necessary. If we just ask for a
model estimator that shows good predictive properties for all regimes
of the
data, then we are aiming for a simple linear model that is reliable
for all possible subsets of the data. We propose the concept of
``maximin effects''
and a suitable estimator and look at its prediction accuracy from a
theoretical point
of view in a mixture model with known or unknown
group structure. Under certain circumstances the estimator can be
computed orders of magnitudes faster than standard penalized
regression estimators, making computations on large-scale data
feasible. Empirical examples complement the novel methodology and theory.
\end{abstract}

%
\end{frontmatter}

\section{Introduction}
``Big data'' often refers to datasets that are large in different ways:
there can be many observations, many variables or both, and the
size can be measured against some historical standard or against
available computational resources (e.g., the data might be too
large to fit into memory). Data can also come from different sources,
have inhomogeneities and might have to be processed in a streaming
fashion. Here, we want to take a look at one specific aspect of ``big
data,'' the effect of inhomogeneities in the data in regression
modeling. Specifically the question whether one is able to extract
(in a computationally feasible way) a model that works for data that
come from different time-regimes or that, more generally, have
different underlying distributions.

From a perhaps slightly naive statistical point of view, a situation
where we face computational challenges due to a large number of
homogeneous observations in a database is not problematic. We can
simply discard most of the
observations and retain sufficiently many observations, chosen at
random, to guarantee
good predictive accuracy. The exact number of observations we have to
retain will be a function of the desired
predictive accuracy, the number of
variables and the noise level. Keeping tens of thousands of
observations will be sufficient for most practical purposes. Most
estimators can easily deal with datasets of this size.

However, many large-scale datasets do not fit neatly into the standard
framework of a single underlying model observed with independent and
identically distributed errors. There are likely to be outliers in the
data, and the truth might better be approximated with a mixture of
models than a single one and underlying distributions of the variables
might shift over time
[\citet{hand2006classifier}].
There has been a lot of work on various aspects of these issues. While
we cannot provide an even approximately complete overview, some of the
major themes can be found in work on robust estimation [\citet{hampel1986robust},
\citeauthor{huber1964robust} (\citeyear{huber1964robust,huber73robust})], time-varying
coefficients models
[\citet{hastie1993varying,fan1999statistical,cai2000efficient}], mixture models
[\citet{aitkin1985estimation,mclachlan2004finite,figueiredo2002unsupervised}]
and change-point estimation [\citet{carlstein1994change}].
In a high-dimensional
regression dataset, \citet{stadler2010} showed evidence for the
presence of
multiple components that can be exploited for variable selection.
Mixed- and random-effects models
[\citet{pinheiro2000linear,mcculloch2006generalized}] are related but do
not have an observation-specific random effect.
Varying-coefficient models seem particularly attractive to capture
shifts in underlying distributions if the data are recorded
chronologically, and the approach has been extended to cope with more
general estimation problems, including estimation of time-varying
graphs [\citet{kolar2010estimating}]. Mixture
models, on the other hand, do not assume such a structure and try to
infer the hidden states of the mixture class membership by using, for
example, the EM-algorithm [\citet{dempster1977maximum}] or related approaches.

%
\begin{figure}

\includegraphics{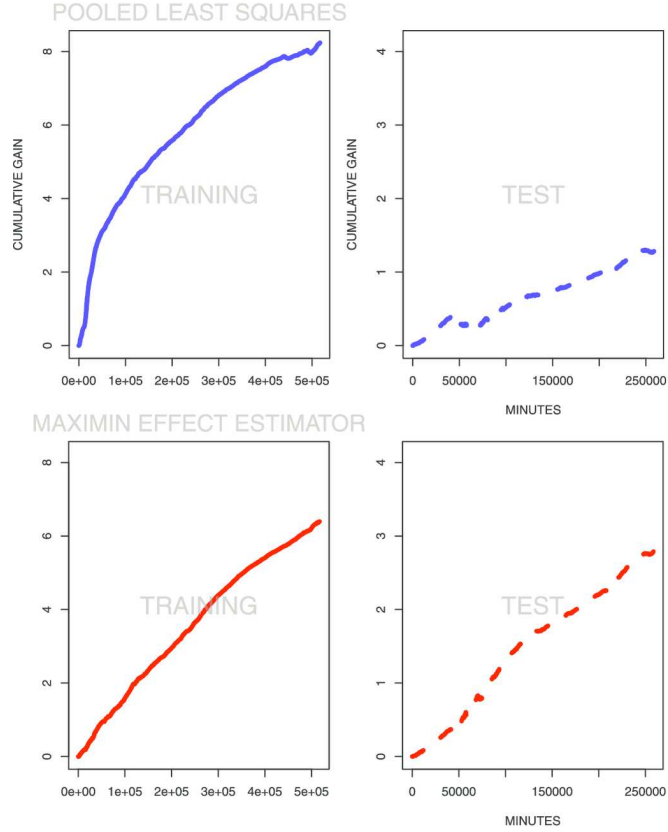}

\caption{Forecasting minute log-returns of the
Euro--Dollar exchange rate with a least squares linear model fit over the
pooled training data and the proposed maximin effects
estimator. The panels show the cumulative cross products as in
(\protect\ref{eqcum}) for the least squares estimator and the maximin
effect estimator, respectively, where the aspect ratios are chosen
such that the same effect strength will lead to the same slope in
all four panels. There are more than half a million training
observations for a model with just 60 free parameters, and yet the
least squares estimator overfits, which leads to a degradation in
performance on the test data. The performance of the maximin
estimator is more consistent over the training data, which
translates into a better performance on the test data.}\label{fig6E}
\end{figure}

In some applications, trying to infer the full time-varying
coefficients in a model or inferring the hidden states in a mixture
model can be computationally challenging, and success is not always
guaranteed from a statistical point of view. Moreover, we might not be
interested in the hidden states or the exact time-evolution of the
coefficients but rather in a simple model that can work reliably for
all states or times. Our examples will mostly fit into a change-point
model, where the underlying distribution can shift abruptly.

An example is given in Figure~\ref{fig6E}, which is based on
price data of twelve financial future instruments (including foreign
exchange, equity and commodity futures) on time-resolution of minutes
over the course of ten years, after 2000. We use the past 5 minutes of
log-returns of all instruments (i.e., 60 predictor variables) to
forecast with a linear model the log-return of the Euro--Dollar
exchange rate over the next
minute (which is the response variable). Two-thirds of the data are
used for training a least squares estimator, and the cumulative cross products
of this model for the training and test data is shown in the first and
second panel of Figure~\ref{fig6E}, respectively, where the cumulative\vspace*{1pt}
gain up to time $1\le t\le n $ is for response values $Y_i$,
$i=1,\ldots,n$ and predictions $\hat{Y}_i$, $i=1,\ldots,n$ (both are
assumed to mean-centered and predictions are normalized to have a
second moment of 1) given by
%
\begin{equation}
\label{eqcum} \sum_{i=1}^t (
Y_i \hat{Y}_i ).
\end{equation}
The training data show that the model works very well at the
beginning of the training period but then tails off, and performance on
the test data is much worse than on the training period, even though
there are more than half a million observations to fit a model with 60
parameters. In contrast, the third and fourth panel show the
cumulative cross products of the ``maximin effects'' estimator that we propose.
The least squares estimator is maximizing the
explained variance on the pooled dataset, leading (as in this example)
to periods or groups of observations where the fit is very good, and
others where barely any variance is explained by the fitted values.
In contrast, the ``maximin effects'' estimator maximizes the explained
variance for the worst group of observations, which have been divided
in this example rather arbitrarily into 3 equally large blocks of consecutive
observations; see Section~\ref{sectionestimator} for more
precise definitions regarding a group of observations.
The estimator in the example is computed without
a regularization penalty.
The predictive accuracy is much more constant over time, and
performance on
the test data is in line with performance on the training data, as the
estimator has not been as much influenced by the period at the
beginning of
the training set as the least squares estimator.

We will set notation
and introduce the maximin effects estimator in Section~\ref
{sectionestimator}, while showing some properties for known and unknown
group structure in Section~\ref{sectionproperties}, discussing
computational properties in Section~\ref{sectioncomputational} and
concluding with an example in Section~\ref{sectionnumerical}.

\section{Maximin effects}\label{sectionestimator}
We will first try to give a suitable and intuitive definition of
maximin effects in mixture models or varying-coefficient models, while
introducing the
maximin effects estimator thereafter.

While we focus exclusively on regression here for ease of exposition,
the same approach can be used, for example, for classification and
graph estimation.

\subsection{Maximin effects for mixture models}
We will work with a mixture model, where for $n$ observations
of a real-valued response $Y_i$ and a $1\times p$ predictor variable
$X_i\in\mathbb{R}^p$ for $i=1,\ldots,n$,
%
\begin{equation}
\label{eqmix} Y_i = X_i B_i +
\varepsilon_i\qquad\mbox{where } B_i\in
\mathbb{R}^p\mbox{ and } B_i \sim F_B
\end{equation}
for some unknown
distribution $F_B$, either discrete or continuous. We also use the
standard notation with the $n \times1$ response vector $Y$, the $n
\times p$ design matrix $X$ and the $n \times1$ error vector
$\varepsilon$. The predictor variables $X_i$
are random and independent, and the noise
$\varepsilon_1,\ldots,\varepsilon_n$ fulfils
$E(\varepsilon^t X)=0$. Furthermore, the coefficients
$B_i$ are independent from the $X_i$, $i=1,\ldots,n$. Independent noise
is an example, but some
dependencies between noise contributions are also possible in this
framework, for example, if the observations have a time-ordering.
The inhomogeneity of
the data is thus solely caused by the variation of the regression
coefficients among the sample points with indices $i=1,\ldots,n$. We
do not necessarily assume
that the $B_i$, $i=1,\ldots,n$
are independent. They can be organized in known or unknown groups.
The following examples indicate the scope of the model: if $F_B$
has point masses at a finite set of points, we are in the setting of
classical finite mixture models, where $B$ can take one of a finite
number of
values. In another scenario, realizations $B_i$ are
positively correlated over time if the observations are
ordered in some chronological order, creating a smoothly varying effect
over time. In the latter example, the model behaves more like a
varying-coefficient model [\citet{hastie1993varying}]. A shift in the
distribution of the predictor variables could conceivably be handled in a
very similar manner. As a final example, the realizations $B_i$
are most often the same, but a small fraction takes other values which
can be viewed as outliers or contaminations.

We always assume that the random $X_i$ are identically distributed
from a
distribution with population Gram matrix $\Sigma$. For a fixed
regression coefficient
$b\in \operatorname{support}(F_B)\subseteq\mathbb{R}^p$,
we can define two different optimality criteria:
$R_{\beta;b}$ is the variance of the residuals in absence of
additional
errors on the observations, while
$V_{\beta;b}$ is the explained variance of predictions with
$\beta\in\mathbb{R}^p$:
%
\begin{eqnarray}
\label{eqR} R_{\beta;b} & =& E \bigl(\llVert Xb-X\beta\rrVert
_2^2/n \bigr) = b^t \Sigma b - 2
\beta^t \Sigma b+\beta^t \Sigma\beta,
\\
\label{eqV} V_{\beta;b} & =& 2 \beta^t \Sigma b -
\beta^t \Sigma\beta.
\end{eqnarray}
Alternative expressions for~(\ref{eqR}) and (\ref{eqV}) under the condition
$E(\varepsilon^t X)=0$ are
\begin{eqnarray*}
R_{\beta;b} & =& E_{Y,X} \bigl( \llVert Y-X\beta\rrVert
_2^2/n \bigr) -E \bigl(\llVert\varepsilon\rrVert
_2^2/n \bigr),
\\
V_{\beta;b} & =& E \bigl(\llVert Y\rrVert_2^2/n
\bigr) - E_{Y,X} \bigl( \llVert Y-X\beta\rrVert_2^2/n
\bigr).
\end{eqnarray*}

If we want to find a single $p$-dimensional regression coefficient
that works optimally on average over $B\sim F_B$, the optimal choice
are the
pooled coefficients
%
\begin{eqnarray}
\label{eqbetapool} b_{\mathrm{pool}} & =& \mathop{\operatorname{argmin}}_\beta E_B(
-V_{\beta;B} ) = \mathop{\operatorname{argmin}}_\beta E_B(
R_{\beta;B} ),
\end{eqnarray}
where the expectation is with respect to $B\sim F_B$.
Note that in this case it is inconsequential for the pooled estimator,
whether we
minimize the residuals or maximize the explained variance.

If we are looking for effects that guarantee a good performance
throughout all possible parameter values, in analogy to decision
theoretic consideration [\citet{wald1945statistical}], two possible
definitions of effects are
%
\begin{eqnarray}
\label{eqbetastarR} b_{\mathrm{pred\mbox{-}maximin}} &=& \mathop{\operatorname{argmin}}_\beta\max
_{b \in F} R_{\beta;b},
\\
b_{\mathrm{maximin}} &=& \mathop{\operatorname{argmin}}_\beta\max
_{b\in F} ( R_{\beta;b} - R_{0;b} )
\nonumber\\[-8pt]\label{eqbetastar} \\[-8pt]\nonumber
&=& \mathop{\operatorname{argmin}}_\beta\max_{b \in F} (-
V_{\beta;b}) = \mathop{\operatorname{argmax}}_\beta\min_{b \in F} (
V_{\beta;b}),
\nonumber
\end{eqnarray}
where $F=\operatorname{support}(F_B)$. Alternatively, $F$ could be the smallest
region such that $P(B \in F) \ge1-\alpha$ for some
$\alpha\in(0,1]$, guaranteeing success for a large fraction of
randomly chosen coefficient values.

Two comments are in order regarding the definition of maximin
effects.
First, the effects are optimizing for the worst-case scenario for $b\in
F$. To be more precise, if we view future samples of $B$ to be allowed
to be chosen by an
adversarial player, the \emph{maximin effects} are then of a minimax
regret form as they optimize the objective function (explained
variance) under the
most adversarial scenario. Minimax regret strategies have also been
explored in game theoretical aspects of decision theory and machine
learning, for example, in
\citet{cesa2006prediction} and \citet{zinkevich2007regret}, and
bandit-type problems and on-line decision
problems,
[e.g., \citet{lai1985asymptotically,foster1999regret,auer2002finite,bartlett2008high,audibert2013regret}].
We do not allow in our framework any choice about which distribution
we sample from, contrary to bandit-type problems. Related to our
setting is a paper by \citet{eldar2004linear}, who propose a
linear minimax
regret estimator which can be computed with convex optimization. Their
estimator is optimizing a mean-squared error loss subject
to various source uncertainties in the data. Our set-up is
conceptually perhaps closest to the minimax framework in robust statistics
[\citet{huber1964robust}].
However, we consider much more general situations than contaminated
samples with a fraction of outliers: as discussed in Section~\ref
{secrobust}, the latter fits into our framework as well.

Second, we have shown two different objectives (minimizing residual
variance $R_{\beta;b}$
and maximizing explained variance $V_{\beta;b}$) that yield two
different minimax-regret
estimators ($b_{\mathrm{pred\mbox{-}maximin}} $ and $b_{\mathrm{maximin}} $).
Using the first choice of minimizing residual variance has the main
drawback that it is nonrobust when sampling regression coefficients:
adding a small point mass to $F_B$ can change the effects
$b_{\mathrm{pred\mbox{-}maximin}} $ drastically. The same is true for the pooled
effects (\ref{eqbetapool}).
Explained variance $V_{\beta;b}$ is expressed as residual variance if measured
against the baseline of residual variance $R_{0,b}$ under a constant 0
prediction
(we assume a mean-centered response
throughout). This baseline is often appealing in practice as we would
like to avoid doing worse than a constant prediction.
Moreover, assume we choose instead a baseline of residual variance
$R_{b_{\mathrm{base}},b}$ for any vector $b_{\mathrm{base}}$ in the convex hull of the
support of $F$ (such as $b_{\mathrm{pool}} $ or $b_{\mathrm{pred\mbox{-}maximin}} $). Whichever
vector $\beta\neq b_{\mathrm{base}}$ we choose in this scenario, we
cannot avoid doing worse than the baseline for some values of $b\in F$
with the consequence that the
problem will become trivial, as the optimal value under the most
adversarial scenario can then always be achieved by a vanishing
coefficient vector (thus keeping the baseline solution). Theorem~\ref
{theoremmaximin} will provide a justification
for this statement: once we shift the problem by the nonzero
baseline $b_{\mathrm{base}}$,
the vector $b_{\mathrm{base}}$ will sit at the origin, and it will be a part of
the convex hull of the equally shifted support of $F$, thus leading to
a vanishing maximin-effect.

%
\begin{figure}

\includegraphics{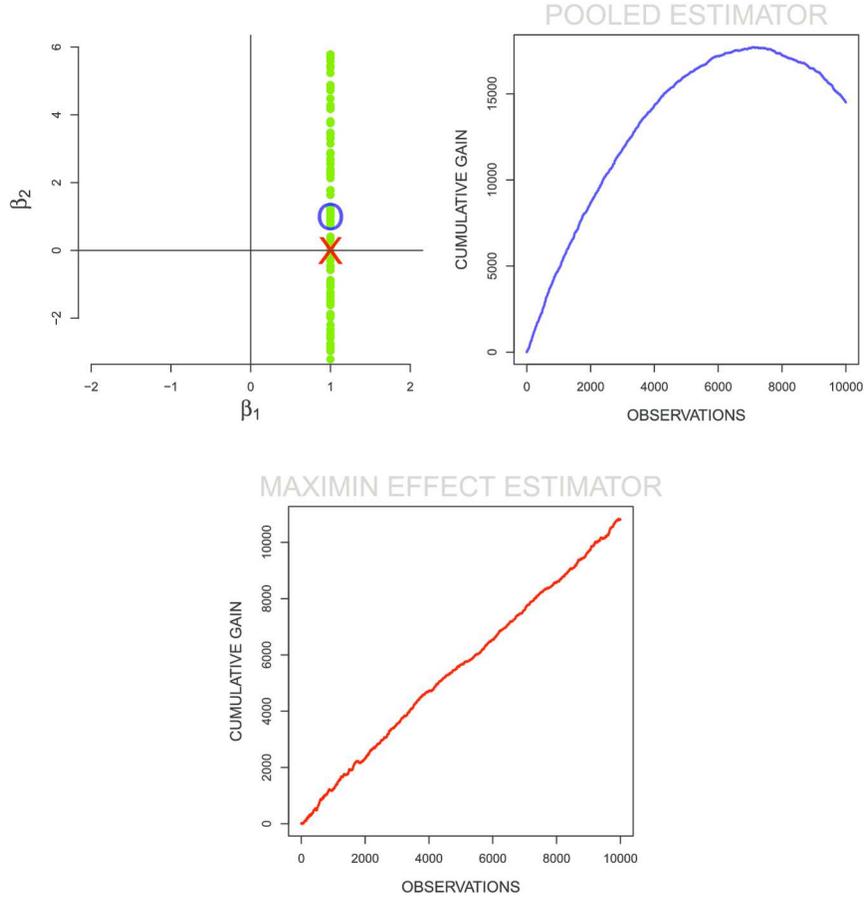}

\caption{An illustration of the difference
between the pooled effects $b_{\mathrm{pool}}$ in (\protect\ref{eqbetapool})
and the maximin effects
$b_{\mathrm{maximin}}$ in (\protect\ref{eqbetastar}). First panel: the green dots
indicate the values the
random coefficient takes on with a maximin first component and a
variable second component. The blue circle indicates the location
of the pooled effects $b_{\mathrm{pool}}$, while the red dot marks
the location of the maximin effects $b_{\mathrm{maximin}}$. Second and third panel: the
cumulative cross product~(\protect\ref{eqcum}) for the pooled and
maximin estimator,
respectively. While the pooled estimator achieves a better overall
fit, it does so at the cost of a highly variable performance.}\label{figsim}
\end{figure}

A simple two-dimensional example of ``maximin effects'' is shown in
Figure~\ref{figsim}. The coefficients are
chosen as $B=(1,\eta)$, where $\eta$ varies uniformly in $[-4,6]$. The two
random predictor variables are chosen independently with a standard
normal distribution. The pooled
estimator (\ref{eqbetapool}) is marked with a blue circle in the first
panel of Figure~\ref{figsim}, and the corresponding cumulative cross
product in~(\ref{eqcum}) is
shown in the second panel if observations are ordered such that
$\eta$ decreases monotonically from $6$ to $-$4. The
pooled effects (\ref{eqbetapool}) maximizes the overall explained
variance but suffers
as $\eta$ takes on negative values. Predictions in
this range are even negatively correlated with the responses, as can be
seen by the negative slope of the cumulative cross product in the figure
toward the right half of the observations. The effect is perhaps more drastic
than in the real-data example in Figure~\ref{fig6E} but of a similar
nature.
The maximin effects $b_{\mathrm{maximin}}$ in contrast take a nonzero value
only for the first variables, where the effect is constant. This
yields a constant explained variance throughout the whole parameter
range, as shown in the third panel of Figure~\ref{figsim}.

If we are in a classical regression model with a fixed regression
coefficient vector, then $F_B$ has just a point mass at some
$b\in\mathbb{R}^p$, and (\ref{eqbetapool}), (\ref{eqbetastarR}) and
(\ref{eqbetastar}) will coincide.
The vector $b_{\mathrm{maximin}}$ is maximizing the explained variance under the
most adverse
realization of the random regression coefficient. The value 0 has a
special status since we define the regret with respect to the 0
regression vector. Effects that can
take opposite signs are set to 0 when using the maximin explained
variance in $b_{\mathrm{maximin}}$ (similar to the value 0 getting special
status when
losing the rotational invariance in coordinate space when replacing a
ridge penalty by a Lasso-type penalty).

The maximin effects have an interesting characterization.

\begin{theorem}\label{theoremmaximin}
Assume that the predictor variables are chosen randomly from a design
with full-rank
population Gram matrix $\Sigma$. Let $H$ be the convex hull of the
support $F$ of $F_B$.
The maximin-effect (\ref{eqbetastar}) is then given by
\[
b_{\mathrm{maximin}} = \mathop{\operatorname{argmin}}_{\gamma\in H} \gamma^t \Sigma
\gamma.
\]
In particular, if $0\in H$, then $b_{\mathrm{maximin}}\equiv0$.
\end{theorem}

\begin{figure}

\includegraphics{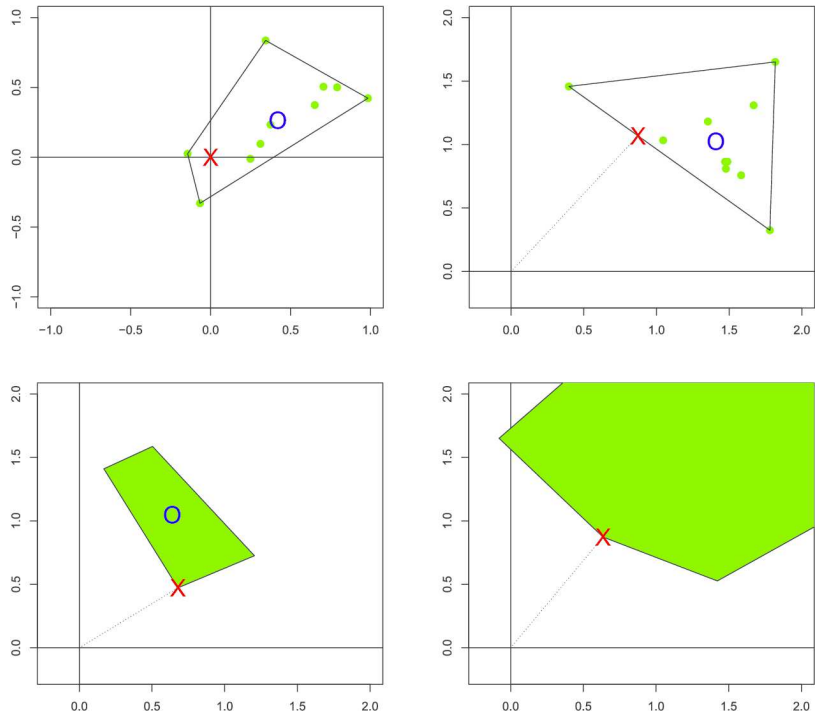}

\caption{Four examples of the support of
$F_B$ (green dots or area), its convex hull (black lines), the pooled
effects (blue
circle) and the maximin effects (red cross). The maximin effects are
the closest point to the origin in the convex hull of the support of
$F_B$ in the distance measure of Theorem~\protect\ref{theoremmaximin}.
In the
example on the first panel, the origin is contained in the
convex hull, and the maximin effects thus vanish. In the second example the
maximin effects rest
on the convex hull of the support, but are
not equal to zero. In both examples, the maximin effects are not part of
the support itself.
The third example shows a continuous support of~$F_B$, while the last
example has unbounded support of~$F_B$. In both of these examples, the
maximin effects are identical to a corner point of the support, but
generalizations to the edge of the support are easily possible as well.
In the last example, the pooled effects are thus infinite
whereas the maximin effects have a robustness property by staying at
the closest point to the origin.}\label{figgeome}
\end{figure}

A proof is given in the \hyperref[append]{Appendix}.
The maximin effects parameter is thus the one that is closest to the origin
in the convex hull spanned by the support of $F_B$. In a classical
regression setting with fixed regression coefficients $\beta^*$, $F_B$
just has
a point-mass at a $\beta^*$, and the maximin effect will, by
Theorem~\ref{theoremmaximin}, be identical to
$\beta^*$. Figures~\ref{fig6E} and \ref{figlasso} show examples of datasets
with an interesting nonzero solution.

If the origin is included in the convex hull of all
coefficients, the best lower bound that can be guaranteed is 0, and the
maximin effects are consequently vanishing identically in this
scenario.
If the maximin effects vanish, a standard regression analysis
will typically be misleading since the inner product between any estimated
vector and true effects in $F_B$ can take an arbitrary sign,
depending on which effect in $F_B$ is currently active. A vanishing
maximin effect is thus a warning sign that standard regression
analysis might not be appropriate.

Four examples of maximin effects are shown in Figure~\ref{figgeome},
comparing the
pooled and the maximin effects. For $b_{\mathrm{pred\mbox{-}maximin}}$, as defined in
(\ref{eqbetastarR}), there is no equivalent characterization as in
Theorem~\ref{theoremmaximin}, as the value 0 has no special status.

It is noteworthy that the maximin effects are robust in the following
sense: if we add new points to the support of $F_B$, we will always
maintain or lower the distance to 0 as in Theorem~\ref{theoremmaximin}. In
the most extreme case, adding contaminations to the support of $F_B$
will either leave the maximin effects unchanged or shrink the maximin
effects toward 0.
This is a direct consequence of Theorem~\ref{theoremmaximin}.

We can also characterize the maximin effects in yet another way, using
Theorem~\ref{theoremmaximin}. Define the predictions and
residuals as
\[
\operatorname{Pred}_\beta:= X\beta\quad\mbox{and}\quad
\operatorname{Res}_{b,\beta}:= Xb - X\beta.
\]
The maximin effects are then, using Theorem~\ref{theoremmaximin},
characterized as the
effects that maximize the norm of the predictions, subject to the
constraint that the inner product between the predictions and the
residuals is nonnegative for all possible $b\in F$,
%
\begin{equation}
\label{eqmax2view}
b_{\mathrm{maximin}} = \mathop{\operatorname{argmax}}_\beta E\bigl
(\llVert
\operatorname{Pred}_\beta\rrVert_2^2\bigr)\qquad\mbox{such that } \min_{b\in F} E( \operatorname{Pred}_\beta,
\operatorname{Res}_{b,\beta}) \ge0.\hspace*{-20pt}
\end{equation}
If $B$ just takes a deterministic fixed value $\beta^*$, we recover of
course $b_{\mathrm{maximin}} =\beta^*$. The constraint in the optimization
above requires that the predictions are never
negatively correlated with the residuals if $b$ can vary in $F$. The
maximin predictions
are in this sense the maximal predictions that are still
``conservative'' in the sense that they can potentially
``under-explain'' a
signal but can never be negatively correlated with the residuals.

In summary, if we want
to maximize the explained variance if an adversary is allowed to pick a
regression vector
$b\in F$ or if test data are not expected to come from the same
distribution as the training data with respect to the random
coefficients, then estimating the maximin coefficients (\ref{eqbetastar})
seems a useful choice.

\subsection{Maximin effects estimator}
We introduce the maximin effects estimator first for data where we know
a group-structure of the observations in the sense that within each
group of observations the regression coefficient has a fixed (but
unknown) value, which varies between groups.

To be more precise, suppose there are $G$ groups of observations
$g=1,\ldots,G$, and each group has $n_g$ samples. The indices
belonging to a group are denoted by $\Ig\subset
\{1,\ldots,n\}$ for all groups $g=1,\ldots,G$. Let $X_{\Ig}=X_g$ denote the
$n_g \times p$-dimensional submatrix of $X$ that corresponds to choosing
the rows in $\Ig$ and likewise for $Y_g=Y_{\Ig}$ and
$\varepsilon_{\Ig}=\varepsilon_g$. If we are in situation where we
know that the random coefficient is fixed at $b_g$ within a group, then
\[
Y_g = X_g b_g + \varepsilon_g,
\qquad g=1,\ldots,G.
\]
Let $\hat{\Sigma}_g= n_g ^{-1} ({X}_g^t X_g)$.
The empirical counterpart to (\ref{eqV}) is the explained variance in
group $g$,
\[
\hat{V}^g_{\beta}:= \frac{2}{n_g} \beta^t
X_g^t Y_g- \beta^t \hat{
\Sigma}_g \beta.
\]

A natural estimator for a sparse, consistent signal $b_{\mathrm{maximin}}$ is
then a penalized version of the empirical minimizer. For $q\in[1,2]$,
%
\begin{equation}
\label{eqestlambda} \hat{\beta}^\lambda= \mathop{\operatorname{argmin}}_{\beta
\in\mathbb{R}^p } L(\beta)
+ \lambda\llVert\beta\rrVert_q\qquad\mbox{where } L(\beta)=\max
_{g=1,\ldots,G} \bigl(-\hat{V}^g_{\beta}\bigr).
\end{equation}
For $G=1$, the loss function $L(\beta)$ is identical to quadratic
loss up to a constant. For $G>1$, however, the loss function will be
different from quadratic loss.
If $p\ll\min_g n_g$, one can use the unpenalized version
($\lambda=0$). In the general case, the two most interesting choices
for the penalty are $q=1$,
making the estimation lasso-like
[\citet{tibshirani96regression,chen01atomic}], and $q=2$ for a ridge-type
estimation [\citet{hoerl1970ridge}]. An equivalent version is given by
the constrained optimization,
%
\begin{equation}
\label{eqest} \hat{\beta}^\kappa= \mathop{\operatorname{argmin}}_{\beta
\in\mathbb{R}^p } \max
_{g=1,\ldots,G} \bigl(-\hat{V}^g_{\beta}\bigr)\qquad
\mbox{such that } \llVert\beta\rrVert_q \le\kappa,
\end{equation}
and we will mostly use the constrained version for the theoretical results.
In practice the two versions can be used interchangeably, and the
penalty parameter can be chosen by
cross-validation, using hold-out samples for each unknown group and
choosing the penalty parameter that maximizes the minimally explained
variance on the hold-out samples from all groups.

The objective function in~(\ref{eqestlambda}) or (\ref{eqest}) is
convex in its argument and
can thus relatively easily be optimized; we will return to this
issue later in Section~\ref{sectioncomputational}.

\subsection{Maximin effects estimator for unknown groups}
In some applications there are no a priori known groups on which the
realized value of the regression coefficients shows little or no
variation. However, if the observations have, for example, a time ordering,
and the effects
are changing smoothly over time, we would suspect that taking blocks
of consecutive observations would result in little variability of the
realized coefficients within groups.

For some datasets, the groups are entirely unknown; see
\citet{stadler2010} for an example. We propose in these cases to take
$G$ groups of~$m$ observations, where $m$ is approximately of size
$n/G$ (modulo rounding to the next integer) if we sample without
replacement. Alternatively, we can sample $G$ groups with $m$
observations each with replacement such that typically $Gm>n$.

For both cases mentioned above, once we have constructed the $G$
groups, we use the same estimator as in (\ref{eqestlambda}) or
(\ref{eqest}). We discuss in Sections~\ref
{subsectionrecoverypareto}--\ref{subsectionpareto} the validity of
the procedure based on such estimated groups.

\section{Properties}\label{sectionproperties}

The statistical properties of the lasso-type maximin effects estimator
(\ref{eqest})
with the $\ell_1$-norm constraint
will be examined first for the case of known
groups in the observations and later be extended to the considerably
more involved case (both
from a theoretical and practical perspective) of unknown groups,
either capturing smooth varying effects (over time) or more generally
without such a (time) structure.

\subsection{Known groups} \label{sectionknown}
Here we show a
result for the lasso-type maximin effect
estimator~(\ref{eqest}) for known groups of observations.

Specifically, there are $G$ groups, and for simplicity, we assume that each
group has
$n_g\equiv n/G$ observations (without this assumption, we need to
replace in the results below $n/G$ by $\min_g n_g$). In each group, the
explanatory variables are chosen randomly with Gram matrix $\Sigma$,
yielding design matrices $X_g$, $g=1,\ldots,G$. In each group,
\[
Y_g = X_g b_g + \varepsilon_g
\qquad\mbox{for }g=1,\ldots,G,
\]
for coefficients $b_g\in\mathbb{R}^p$ that are fixed in each group
but can vary between groups. The maximin estimator is then the set of
coefficients that work optimally across all groups in the sense of~(\ref
{eqbetastar}).

With estimator (\ref{eqest}), we are now trying to maximize the explained
variance in all groups.

%
\begin{theorem} \label{theorembasic}
Let $D$ be the maximal difference between the empirical Gram matrix $
\hat{\Sigma}_g$ and population Gram matrix $\Sigma$, $D = \max
_g\llVert \hat{\Sigma}_g-\Sigma\rrVert _\infty$.
If $\kappa\ge\max_g\llVert b_g\rrVert _1$, then
%
\begin{equation}
\label{eqknownres} \min_{b\in F} V_{\hat{\beta}^\kappa;b} \ge V^* -6D
\kappa^2 -\frac{4}{n/G} \max_g \bigl\llVert
X_g^t \varepsilon_g\bigr\rrVert
_\infty\kappa,
\end{equation}
where $V^*$ is the optimal value that can be attained,
%
\begin{equation}
\label{eqVstar} V^* = \max_\beta\min_{b\in F}
V_{\beta;b}.
\end{equation}
\end{theorem}

A proof is given in the \hyperref[append]{Appendix}. For $D=0$ (if the population and
empirical versions of the Gram matrices are identical, as happens under
fixed design; see a more detailed comment below), the estimator thus
reaches the
optimal value less a term
\[
\max_g \frac{4}{n/G} \bigl\llVert X_g^t
\varepsilon_g\bigr\rrVert_\infty\max_g
\llVert b_g\rrVert_1,
\]
which is a similar result to that of standard lasso estimation
[see, e.g., \citet{buhlmann2011statistics}], except that the first term
$4 (n/G)^{-1} \times\break \max_g \llVert X_g^t \varepsilon_g\rrVert _\infty$ will
increase when the
number $G$ of groups grows larger, which is the price we have to pay
for estimating the maximin effects (\ref{eqbetastar}) instead of the
pooled effect (\ref{eqbetapool}). On the other hand, the error is
just a function of the noise $\varepsilon$ and not influenced by
the variability of $b_g$ across groups, whereas standard lasso-type
estimation of the pooled effect (\ref{eqbetapool}) would suffer if
the variability of $b_g$ is high across groups.

We note that one can also derive a similar bound if the Gram matrix of
the predictors is allowed to depend on the group. In particular, we
can have a fixed design in each group. In this case a corresponding result
holds true with $D=0$. If the design is random as in Theorem~\ref
{theorembasic}, we have an additional
term in the bound that is proportional to $D$ times the squared
$\ell_1$-norm of $b_{\mathrm{maximin}}$ and $b_g$, $g=1,\ldots,G$. A more
careful analysis for the special case of Gaussian random design
[\citet{raskutti2010restricted}] could render the bound again linear in
$\kappa$, with more general design treated in \citet
{guillaume2014compressed}.

The two terms that are relevant for the rate are thus $D$ and $\max
_g\llVert X_g^t \varepsilon_g\rrVert _\infty$.
To give a simple bound for $D$ if all predictor variables are drawn
from the same
population with Gram matrix $\Sigma$, we can, for example, get the following:

\begin{lemma}\label{lemmadeviation}
If the predictor variables are chosen i.i.d. from a distribution with
Gram matrix
$\Sigma$ and $\llVert X_i\rrVert _\infty\le M $ for $i=1,\ldots,n$,
then for any $\alpha\in
(0,1)$, with probability at
least $1-\alpha$,
%
\begin{equation}
\label{eqdeviation} D^2 \le\frac{2M^2}{n/G} \log\biggl(
\frac{2p^2G}{\alpha} \biggr).
\end{equation}
\end{lemma}

The proof follows directly from Hoeffding's inequality, combined with
a union bound over both the~$p^2$ entries in each empirical Gram
matrix and the number $G$ of groups. Of course, a similar bound could
be derived for a Gaussian or sub-Gaussian distribution of the
explanatory variables.

If we additionally make a distributional assumption for the independent
noise to control the term $\max_g\llVert X_g^t \varepsilon_g\rrVert
_\infty$, we get a simple bound for the estimator in (\ref{eqest}).

\begin{corollary}\label{corfixed}
Assume that the predictor variables are chosen i.i.d. from a
distribution\vspace*{1pt} with Gram matrix $\Sigma$ and $\llVert X\rrVert _\infty
\le1$. If the
errors
$\varepsilon_i$, $i=1,\ldots,n$
have a i.i.d. Gaussian distribution $\mathcal{N}(0,\sigma^2)$, then if
$\kappa\ge\max_g \llVert b_g\rrVert _1$,
with probability at least $1-\alpha$,
\[
\min_{b\in F} V_{\hat{\beta}^\kappa;b} \ge V^* - \frac{1}{\sqrt{n/G}}
\biggl[ 6 \sqrt{2\log\biggl( \frac{2p^2G}{\alpha}\biggr)} \kappa^2 + 4
\sigma\sqrt{ 2\log\biggl(\frac{2pG}{\sqrt{2\pi} \alpha}\biggr)} \kappa \biggr].
\]
\end{corollary}

The error features the same $(n/G)^{-1}$ rate as lasso estimation on a
single block
of homogeneous data with $n/G$ observations of a fixed signal.
The success hinges obviously on the sparsity of the maximin
solution.
The bound becomes less tight when $\kappa$ grows. Observe that
$\kappa$ is constrained from below by the sparsity of the regression
coefficients. The problem thus becomes easier for sparse regression
coefficients as one would expect from standard sparse regression
[\citet{buhlmann2011statistics}].

In summary, the maximin effects estimator (\ref{eqest}) is able to
estimate the maximin effects in a dataset with known groups. Note that
an alternative would involve computing the Lasso-type estimator on
each group and then constructing the estimator that yields the best
minimally explained variance across all groups.
In presence of a large number $G$ of groups, the statistical properties
of such a naive alternative procedure are unclear.

\subsection{Unknown groups}

The more difficult scenario is a mixture model, where there is no
a priori known group structure for the observations, and each
observation has its own realized value of the random coefficients.
We assume that each
coefficient $B\in F=\operatorname{support}(F_B)$, where $F$ is compact.

As mentioned previously, for the case of unknown groups, one solution
is to apply estimator~(\ref{eqest}) to chosen groups, which can be
chosen at random in the absence of any group information or in some
way that reflects prior knowledge, for example, in the case coefficients
varying over time.

\subsubsection{Pareto condition}\label{sectionpareto}
We will need to make one main assumption for recovery of the maximin
coefficients for unknown groups, which will be discussed
in a few examples below.

First, we define an essential subset of regression coefficients.

\begin{assumption}[(Essential subset)]\label{AES}
A set $A=\{b_j; b_j\in F\}_{j=1,\ldots,d}$ is called an essential subset
of $F=\operatorname{support}(F_B)$ if the maximin effects for $B\sim\tilde
{F}_B$, where the
support of $\tilde{F}_B$ is $A$, are identical to the
maximin effects as for the original problem with $B\sim F_B$.
\end{assumption}

An essential subset is at most of cardinality $d\le p$ (if $d>p$,
at least one $b_j$ could be removed without changing the point that
is closest to the origin in
the convex hull of these points).

Two examples serve as simple illustrations: if $b_{\mathrm{maximin}} \in F$,
then the smallest essential subset is just
$b_{\mathrm{maximin}}$ itself. If $F$ is discrete, then an essential subset is
always the support of $F$ itself.

We now give the so-called Pareto condition which will be shown to be
sufficient for recovery. For known groups, we do not need the
condition, as it always fulfilled, as in Section~\ref{sectionpareto}.
The condition is meant for cases where the groups are
sampled randomly according to some mechanism, which we discuss with a
few examples and settings in Section~\ref{sectionpareto}.

\begin{assumption}[(Pareto condition)]\label{A1}
Let $I_g\subset\{1,\ldots,n\}$ be the index sets for chosen groups
$g=1,\ldots,G$, and let $B_i$, $i=1,\ldots,n$ be the regression
coefficients at observation $i\in\{1,\ldots,n\}$.
The assumption is that, for $\gamma\in(0,1)$, with probability
$1-\gamma$ with respect to the
random coefficients $B_i$, $i=1,\ldots,n$ and a potentially random
sampling of the groups, there exists an essential subset $A$ of $F$
such that
for each $b\in A$ there exists a group $g\in\{1,\ldots,G\}$ for which
$B_i=b$ for all $i\in I_g$.
\end{assumption}

We call this the Pareto condition since it implies that the maximin
vector is optimal in the sense that making the
performance better in one group will make the performance worse in
another group. The condition requires some of the groups to be
``pure'' in the sense that all observations in the group correspond to
the same realization of the regression
vector.
We emphasize that the Pareto condition is formulated as the
probability of
a certain event: we find this construction simpler than
requiring a random event condition.

The Pareto condition is fulfilled for a few examples which
will be discussed further in Section~\ref{subsectionpareto}, but
the condition is not true in general. Without appropriate structure
(of the type shown in the examples) cases exist where the condition is
violated.

\subsubsection{Recovery assuming the Pareto condition}\label{subsectionrecoverypareto}

Using the Pareto condition,
we get the following theorem for randomly sampled groups in the
estimator (\ref{eqest}):

%
\begin{theorem}\label{theoremrandom} Assume the Pareto condition is
fulfilled, with corresponding probability $1-\gamma$ for some
$\gamma\ge0$. If $X_i$, $i=1,\ldots,n$ are i.i.d. from a
distribution with Gram matrix $\Sigma$
and $\llVert X\rrVert _\infty\le1$ and
$\kappa\ge\max_g\llVert b_g\rrVert _1$, then with
probability at least $1-3\alpha-\gamma$,
%
\begin{eqnarray}\label{eqtheorand}
\min_{b\in F} V_{\hat{\beta}^\kappa;b} &\ge& V^*- \frac{M}{\sqrt{m}}\quad\mbox{and}\nonumber
\\
\bigl(\hat{\beta}^\kappa-b_{\mathrm{maximin}}
\bigr)^t \Sigma\bigl(\hat{\beta}^\kappa-b_{\mathrm{maximin}}
\bigr) &\le&\frac{M}{\sqrt{m}}
\\
\eqntext{\displaystyle\mbox{with } M = 6 \sqrt{2\log\biggl(\frac{2 p^2 G}{\alpha} \biggr)}
\kappa^2 +4 \max_g \frac{1} {\sqrt{m}} \bigl\llVert
X^t_g \varepsilon_g\bigr\rrVert
_\infty\kappa.}
\end{eqnarray}
\end{theorem}

A proof is given in the \hyperref[append]{Appendix}.
If the smallest eigenvalue of the population covariance matrix
$\Sigma$ is bounded from below by some $\lambda_{\min}>0$, then it
follows further that with the same probability as above,
$\llVert \hat{\beta}^\kappa-b_{\mathrm{maximin}}\rrVert _2^2 \le M/ (\sqrt
{m}\lambda_{\min})$.

If the error distribution is Gaussian, we get the following:

\begin{corollary}\label{corrandom}
If the assumptions of Theorem~\ref{theoremrandom} are fulfilled, and
additionally the errors
$\varepsilon_i$, $i=1,\ldots,n$ have a i.i.d. Gaussian distribution
$\mathcal{N}(0,\sigma^2)$, then,
with probability at least $1-3\alpha-\gamma$,
the constant $M$ in Theorem \ref{theoremrandom} can be chosen as
\begin{eqnarray*}
M &=& 6 \sqrt{2\log\biggl(\frac{2 p^2 G}{\alpha} \biggr)}\kappa^2 + 4
\sigma\sqrt{ 2\log\biggl(\frac{2pG}{\sqrt{2\pi} \alpha} \biggr)}\kappa.
\end{eqnarray*}
%
\end{corollary}

This result is a generalization of Corollary~\ref{corfixed}. If we
choose $m=n/G$, we obtain the results of Corollary~\ref{corfixed}.
(Note, however, that we need the Pareto condition for
Corollary~\ref{corrandom} but not Corollary~\ref{corfixed}.)
However, the results also show that we can choose $m$ much larger
than $n/G$ by allowing an observation to appear in multiple groups,
thus lowering the statistical error. Another point of view is that we
keep the sample size in each group fixed but increase the number of
groups $G$, thus increasing the chance that the Pareto condition will be
fulfilled.
We can thus infer the optimal maximin coefficients by randomly sampling
groups and applying the maximin estimator (\ref{eqest}) to these
groups. The success hinges on the sparsity of the coefficients
within the support of the distribution of the random coefficients.

We describe in Section~\ref{sectionnumerical} a cross-validation method
for choosing the number of groups.

\subsubsection{Examples where the Pareto condition is fulfilled}\label{subsectionpareto}

Theorem~\ref{theoremrandom} rests on the Pareto condition. It is
evident that an arbitrary random sampling scheme cannot lead to success
in the
sense of Theorem~\ref{theoremrandom}. The number $G$ of groups, for example,
plays a crucial part. Setting $G=1$ leads just to pooled estimation,
which yields in general a consistent estimator for the pooled
coefficients and can thus not consistently estimate the maximin
coefficients if they differ from the pooled coefficients.

\paragraph*{Fixed groups with fixed design}
The simplest example where the Pareto condition is fulfilled is the
case of known groups, where $B$ takes a single value $b_g$ within each
group $g=1,\ldots,G$. By definition of the maximin coefficients, the
Pareto condition is then fulfilled, and we are back to the setting of
Section~\ref{sectionknown}.

\subsubsection*{Chronological observations with a jump process}
Assume the observations have a time-ordering, and we have a
change-point model. Consider the case where
the support of $F_B$ is finite of cardinality $J$, that is, $F =
\{b_1,\ldots, b_J\}$. Assume that $B$ for the first sample, namely
$B_1$, is chosen uniformly at random
among the $J$ different possibilities. Thereafter, for
$i=2,\ldots,n$ and some $\delta\in(0,1)$
\[
B_i = \cases{ B_{i-1}, &\quad w.p. $1-\delta$,
\cr
b_j, &\quad w.p. $\delta/J$ for all $j=1,\ldots,J$.}
\]
We build $G$ groups of consecutive observations. Below, we will further show
under which conditions on $G$ and $J$ the Pareto condition is
fulfilled with high probability. The Pareto condition is fulfilled if
we have for
each possible value $b_1,\ldots,b_J$ a $g\in\{1,\ldots,G\}$ such
that $B_i=b_j$ for all observations $i$ in the $g$th set. Suppose
we fix $G$ and condition on $B_{i'}=b_j$ for some $j\in\{1,\ldots,J\}$
and some
$i'\in\{1,\ldots,n-2n/G\}$. Let $L$
be the conditional length of the segment of observations $i\ge i'$
where $B_{i}=b_j$. Then
\[
P\biggl(L\ge2\frac{n}{G}\biggr) \ge(1-\delta)^{2n/G} \ge1-
\frac{2n\delta}{G}.
\]
If indeed, $B_{i'}=b_j$ and $L\ge2n/G$, then one block of
observations of length $n/G$ is guaranteed to have exclusively
realizations of $B$ equal to $b_j$.
The probability that there is at least one $i'$ for which
$B_{i'}=b_j$ in $\{1,\ldots,n-2n/G\}$ is greater than
\[
1-\biggl(1-\frac{\delta}{J}\biggr)^{n-2n/G} \ge1-\exp\biggl(-
\frac{\delta}{J}(n-2n/G)\biggr).
\]
Using a union bound over all $J$ distinct values the coefficients can
take, the Pareto condition is fulfilled with corresponding probability
at least
$1-\gamma$ for $\gamma\in(0,1)$ if
\begin{eqnarray*}
G & \ge&4 \frac{n\delta J}{\gamma},
\\
\frac{\delta(n-1)}{J} &\ge&1/ \log\biggl( \frac{2J}{\gamma} \biggr).
\end{eqnarray*}
The second condition states that the number of distinct classes
$J$ cannot be too large. More specifically, $\delta n/J$ is
approximately the average number of contiguous blocks of observations
that have a realization $b_j$ of the random coefficient. The condition
above implies that this average
value needs to be
larger than 1 for the scheme to work (as otherwise a value of the
coefficients might not
be sampled at all).

The first condition is a requirement on the number of groups $G$ one
has to pick. It yields an effective sample size $n/G$ of order
$\delta^{-1}$, which is the order of the length of observations where
the regression coefficient stays constant.

\paragraph*{Contaminated samples and robustness}\label{secrobust}
Assume that the regression coefficients come from a mixture
distribution
%
\begin{equation}
\label{eqcontam} B = \cases{ b^*, &\quad with probability $1-\epsilon$,
\cr
U, &
\quad otherwise,}
\end{equation}
where $U$ follows a distribution $F_U$ such that
%
\begin{equation}
\label{eqexcon} \bigl(u-b^*\bigr)^t \Sigma b \ge0 \qquad\forall u\in
\operatorname{support}(F_U).
\end{equation}
Note that the latter condition implies that $ b_{\mathrm{maximin}}=b^*$.
A fraction $\epsilon$ of samples are contaminated in the
sense that they have a different realized value of $B$.

We build $G$ groups of observations by random sampling. Each group
consists of $m$ samples drawn at random without replacement from all
$n$ observations, and each group is sampled independently. We will
argue under which circumstances the Pareto condition is fulfilled with
high probability.

The
Pareto condition is trivially fulfilled if we have a single group of
observations where all realizations are identical to $b^* = b_{\mathrm{maximin}}$.
Suppose we divide the samples into $G$ groups. Each group contains $m$
observations, drawn at random without replacement from all $n$
observations, independently for each group (and thus, the same
re-sampled data point can occur in several groups).
If for $\gamma\in(0,1)$%
\begin{equation}
\label{eqcondG} G\ge\frac{\log({1}/\gamma)}{\log(1-(1-\epsilon)^m)},
\end{equation}
then we guarantee the Pareto condition will be fulfilled with
corresponding probability at least
$1-\gamma$ with respect to the random sampling of the coefficients and
random sampling of the groups.

There is also a robustness inherent in the procedure.
If sampling~(\ref{eqcontam}) holds without condition
(\ref{eqexcon}), then the samples $U$ can come from an arbitrary
distribution. If condition (\ref{eqcondG}) is fulfilled, then we have
again with probability at least $1-\gamma$ that there is a group where
$B$ is equal to $b^*$ for all observations in the
group. We can then use Theorem~\ref{theoremmaximin} to show robustness
properties of the estimate, as already discussed in the paragraphs
after Theorem~\ref{theoremmaximin}. Adding contaminated samples can
only shrink the maximin effects parameter and the corresponding
estimator toward the origin.
The maximin
effects estimator thus has robustness properties against outliers as
long as at least one group does not contain outliers.

Some more examples are possible to derive, including for continuous
distributions of $B$, but are beyond the scope of this manuscript. The
basic intuition is that the convex hull of the effective coefficients
in each groups needs to approximate the convex hull of the support of
the random coefficients $B$ in order for the Pareto condition to be
fulfilled.

The outliers above are referred to as \textit{b-outliers} in linear
mixed models [\citet{mcculloch2001generalized}]. An interesting question
is whether the method is also robust to outliers in the noise, the
so-called \textit{e-outliers}. If we use a robust version of the explained
variance $V_{\beta;b}$ in the maximim estimator definition, then the
breakdown point of the maximin estimator will at least be identical to
the breakdown point of the robust estimator for the explained
variance. The reason is that the explained variance would have to be
corrupted arbitrarily much
in every of the $G$ groups, requiring in each group $g=1,\ldots,G$ at
least $\lceil \rho n_g \rceil$ corrupted samples
if $\rho\in(0,1)$ is the breakdown point of the robust explained variance
estimator. Hence at least $\sum_{g} \rho n_g=\rho n$ samples would have
to be
corrupted for the modified maximin estimator to take on arbitrarily
large values, and the breakdown point of a robust explained-variance
estimator would thus be inherited by the maximin
estimator.

Before presenting some numerical results, we first discuss now the computational
aspects of the procedure.

\section{Computation}
\label{sectioncomputational}

The objective function of estimator~(\ref{eqestlambda}) is convex, and the
penalty is separable.
Estimator~(\ref{eqestlambda}) or the equivalent constrained
formulation (\ref{eqest}) could thus be computed using
coordinate-wise updates, with a similar strategy as in the ``glmnet''
approach
[\citet{friedman2009glmnet}] to fitting lasso- and ridge-penalized
regression models.
If $p$ and $n$ are large, this becomes
computationally burdensome.
We show two different possibilities.

\subsection{Iteratively reweighted estimation}
The estimation can be
reduced to a series of weighted standard lasso or ridge estimation.
The minimum in~(\ref{eqest}) can be approximated for positive terms by
a sum
%
\begin{equation}
\label{eqestlambdaapprox} \lim_{\zeta\rightarrow0} \Biggl( \sum
_{g=1}^G \bigl( \hat{V}^g_\beta
\bigr) ^\zeta\Biggr) ^{1/\zeta}.
\end{equation}
This leads to a weighted estimation problem, where the weights
are constant in each group, and weights are larger in groups where
the explained variance is still small. For a fixed value of
$\zeta>0$, the solution of (\ref{eqestlambdaapprox}) is (setting
$q=1$ for Lasso-type estimation and $q=2$ for ridge)
%
\begin{equation}
\label{reweighted} \mathop{\operatorname{argmin}}_{\beta\in\mathbb{R}^p} -2 \sum_{i=1}^n
w_i Y_i (X\beta)_i + \sum
_{i=1}^n w_i (X\beta)_i^2
+\lambda\llVert\beta\rrVert_q,
\end{equation}
where the weights $w_i$, $i=1,\ldots,n$ are proportional to
%
\begin{equation}
\label{weights} w_i \propto\bigl(V^{g_i}_{\hat{\beta}}
\bigr)^{\zeta-1},
\end{equation}
where $g_i$ is the group that observation $i$ belongs to. The strategy
is now to alternate between updating the weights in (\ref{weights}),
starting with uniform weights, and computing the solution to
(\ref{reweighted}) for fixed weights.
The solution in the first example in
Figure~\ref{fig6E} has been computed in this way, as a series of
reweighted least squares estimators with $\zeta=0.01$. A few rounds of
the iteration
are typically sufficient, and the computational burden is thus similar
to standard least squares or lasso-type estimation.

\subsection{Computationally efficient solution for maximal penalties}
Computing estimator (\ref{eqestlambda}) by coordinate-wise updates or
by iteratively reweighted penalized estimation requires,
however, that either the design matrix $X$ or the Gram matrices
in all groups are kept in memory.

Another option is to look at the limit of $\hat{\beta}^\lambda$ as
$\lambda\rightarrow
\lambda_{\max}$, where $\lambda_{\max}$ is the supremum of the values
for which the estimator does not vanish identically. In this limit,
\[
\frac{ \hat{\beta}^\lambda} {\llVert \hat{\beta}^\lambda\rrVert _2}
\rightarrow\frac{\hat{\beta} }{\llVert \hat{\beta}\rrVert _2},
\]
where $\hat{\beta}$ is the solution to
%
\begin{equation}
\label{eqmax} \hat{\beta}:= \mathop{\operatorname{argmin}}_{\beta\in\mathbb
{R}^p} \llVert\beta\rrVert
_1\qquad\mbox{such that } \min_{g=1,\ldots,G} \bigl(
\beta^t {X}_g^t Y_g \bigr) \ge1.
\end{equation}
The quadratic term disappears in~(\ref{eqmax}) as it will shrink like
$\kappa^2$ if $\llVert \beta\rrVert _1=\kappa$, whereas the remaining
two terms
(penalty and objective) in the estimator scale linearly with~$\kappa$,
and the constant $\kappa$ thus drops out of the equation or can be
replaced with an arbitrary constant (modulo scaling of the solution)
as $\kappa\rightarrow0$.
The constant 1 is chosen arbitrarily, and choosing a different constant
would just rescale the solution. The estimator $\hat{\beta}$
in~(\ref{eqmax}) can be
computed with linear programming, and most importantly, the data
matrix ${X}_g$ enters only through its inner product with the response
vector $Y_g$, achieving a great reduction in problem size.
Estimator (\ref{eqmax}) still has to be re-scaled for optimal
least squares prediction, but this is just a univariate
optimization. Our only tuning parameter in this case is the
number of groups $G$ to choose (unless they are known), and we can
optimize $G$ by using cross-validation; see the section with numerical
examples for details on how the cross-validation is implemented.

The solution $\hat{\beta}$ in (\ref{eqmax}) selects in general several
variables, not
just a single one as might be expected from the analogous situation
for the standard lasso. For ridge estimation with $q=2$,
estimator~(\ref{eqmax}) would correspond to marginal regression if we
had only a single group, and this behavior has recently been
examined in \citet{genovese2012comparison}. However, the variability
of the inner products in~(\ref{eqmax}) across groups leads to
sometimes appreciably different solutions and has a similar effect as
the quadratic penalty that is written down explicitly in
(\ref{eqestlambda}). We will use this latter estimation technique in
(\ref{eqmax}) for the
following high-dimensional example in Section~\ref{sectionnumerical},
which will also demonstrate the
computational advantages of this type of estimation.

The maximal penalty solution is suitable if either a fast initial
estimator is desirable or if the data are very noisy. In the latter
case the large penalty will be justified not only from a computational
but also from a statistical point of view. The performance of the
maximal penalty estimator should be seen as a lower bound for what is
achievable with a more expensive estimation with a fine-tuned value of
the penalty parameter.

While tight bounds on the worst-case and typical computational
complexity are difficult
to establish, the memory requirements are more immediate. If fitting a
pooled estimation, the required memory is of order $O(\min\{ p^2,
np\})$, as either the whole matrix $X$ or the Gram matrix has to be
held in memory. For standard maximin estimation or estimation of mixture
models with $G$ groups, this is increased to $O(\min\{G p^2,
np\})$ since the Gram matrix has to be stored separately for all $G$
groups. For the maximal penalty estimator (\ref{eqmax}) or its ridge
counter-part, the memory
complexity decreases to $O(pG)$, as one only has to store the $G$
$p$-dimensional cross-products between the predictors and the
responses in each group. The
memory complexity of the maximal-penalty estimator is thus substantially
reduced in the typical scenario where $G\ll\min\{n,p\}$, while the
memory complexity is just marginally increased for the general
case with arbitrary $\lambda$.

\section{Numerical example}\label{sectionnumerical}

The example in Figure~\ref{fig6E} illustrated that overfitting has to
be a
concern even if we have millions of observations at our disposal to
fit quite low-dimensional models with less than one hundred parameters
due to the shifts in the underlying distributions.

Next, we look at an example with millions of variables and thousands of
observations.
\citet{kogan2009predicting} collected a dataset of so-called
{10-\textit{K}} reports from
thousands of publicly traded U.S. companies in the years
1996--2000. For each report, unigrams and bigrams of word frequencies
have been computed and used as predictors for the stock return
volatility in the
twelve-month period after the release of the report, which is here
measured against the baseline of the volatility before the filing of
the report. We use 3000 examples as a training set and the remaining
just over 16,000 examples for testing. We compute both a cross-validated
lasso and ridge estimator with the ``glmnet'' package
[\citet{friedman2009glmnet}] and the estimator (\ref{eqmax}), once for a
fixed number of 3 groups and once for a cross-validated number of
groups, which is explained in the next paragraph. Below, we will further explain the
procedure for cross-validation.

The histograms of%
\begin{equation}
\label{eqexplvar} \sum_{i\in I} Y_i
\hat{Y}_i
\end{equation}
are shown for the various methods, where both $Y$ and $\hat{Y}$ are
standardized to have mean 0 and variance 1. The groups in $I$ are
chosen randomly as 500
observations out of the training or test data. Form
(\ref{eqexplvar}) avoids the choice of the scaling for the estimators
of form (\ref{eqmax}) as the measure is invariant under rescaling
of the predictions.

\begin{sidewaysfigure}

\includegraphics{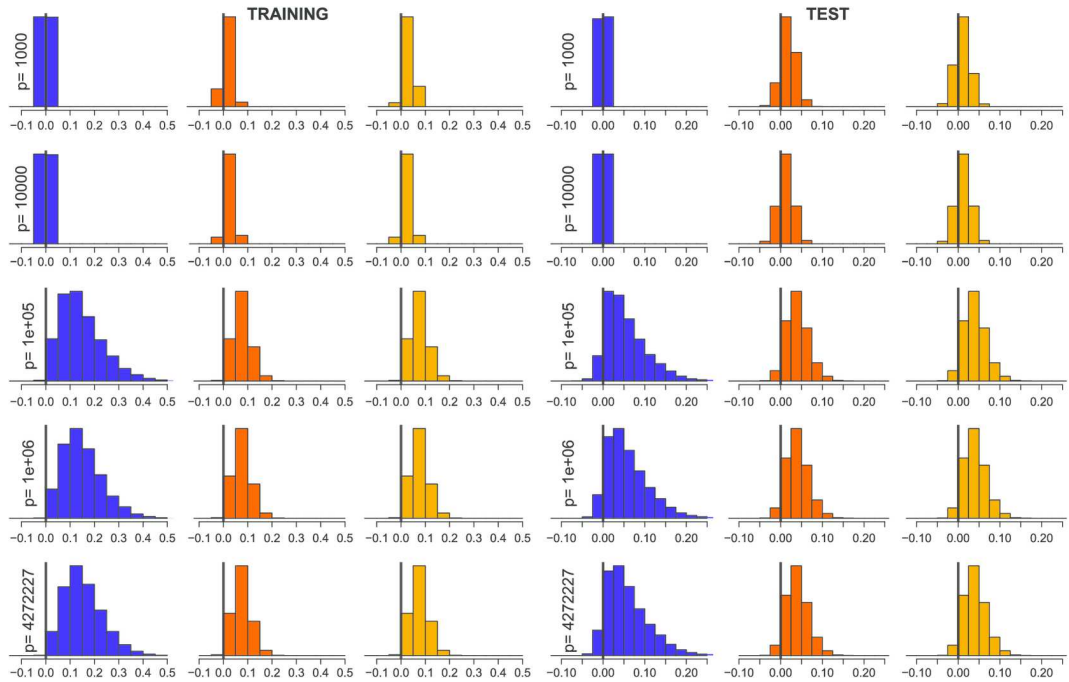}

\caption{The histograms of the cross product~(\protect\ref{eqexplvar})
for lasso-type estimation on the
training data (left three columns) and test data (right three
columns). The predictions are standardized, and the measure is
thus equivalent to explained variance after rescaling of the
predictions. The number of predictor variables is increasing from the
top to the bottom row, from 1000 to 4,272,227.
The three columns in each panel correspond to standard cross-validated lasso
estimation (blue) and the two maximin effects estimators with a
fixed number of $G=3$ groups (red) and a cross-validated choice of
$G$ (orange). In both training and test data, the explained variance
is more variable under the pooled estimation than when estimating the
maximin effects, while the average explained variance is similar. There
is little difference between the estimator with a cross-validated
choice of the number of groups and a fixed number of $G=3$ groups. For
$p\le10^4$, the cross-validated lasso estimator returns an empty
model while the maximin estimation still finds some signal in the data,
even if it is weaker than when using $p\ge10^5$ predictor
variables.}\vspace*{-5pt}\label{figlasso}
\end{sidewaysfigure}

\begin{sidewaysfigure}

\includegraphics{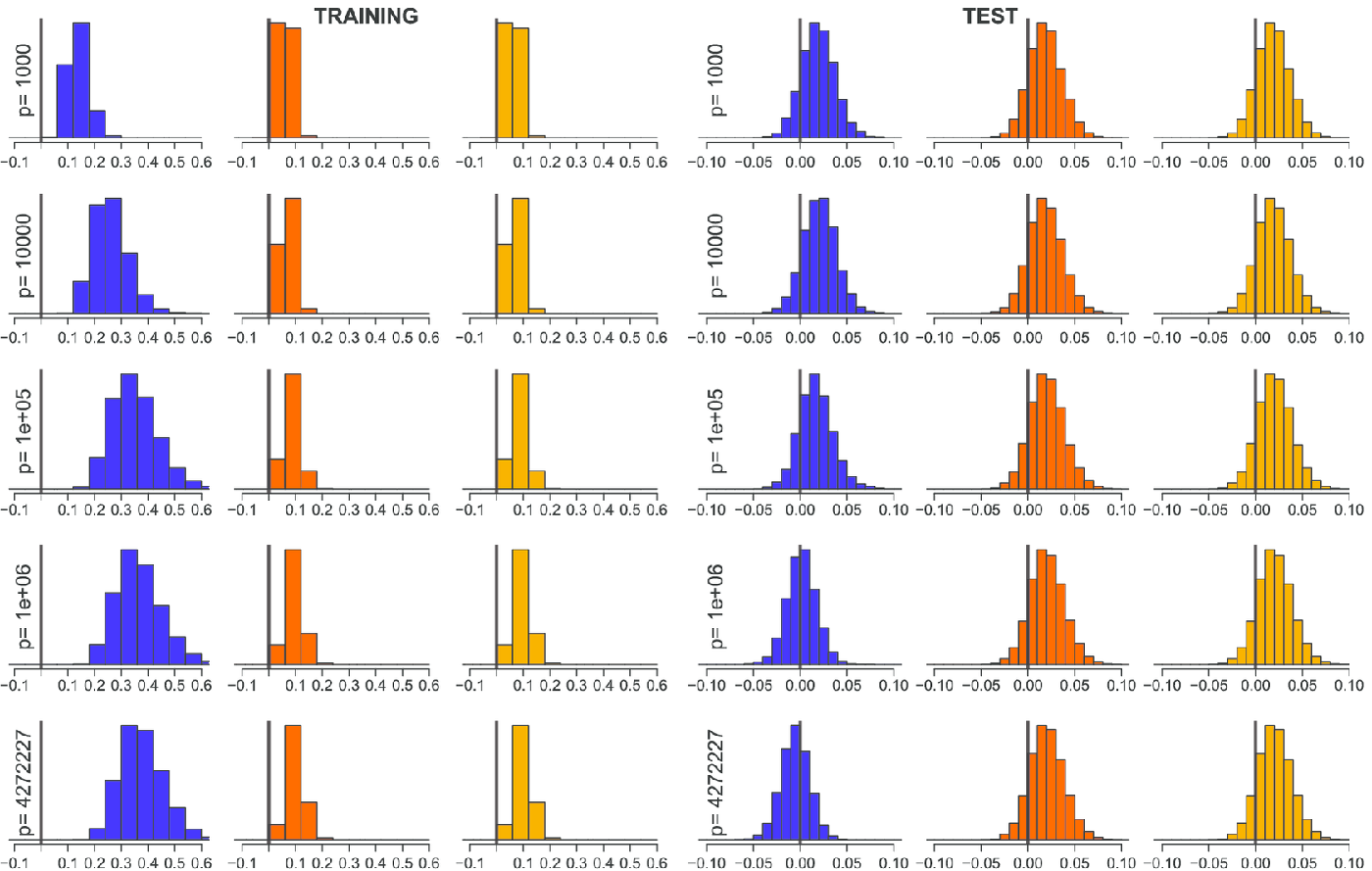}

\caption{The results for ridge regression,
analogous to Figure~\protect\ref{figlasso}. The gap between training and
test performance is much more pronounced for the pooled ridge
estimate (blue) compared to the estimators of maximin effects
(orange and red). Moreover,
the probability of having a subset of observations with very small
(or negative) explained variance is slightly higher for the pooled
estimation. Estimation of the maximin effects was also three orders
of magnitudes faster for the fixed number of groups $G$ (red) than the
pooled estimation (blue).}\label{figridge}
\end{sidewaysfigure}

The results for lasso estimation ($q=1$) are shown in Figure~\ref{figlasso}
and for ridge estimation ($q=2$) in Figure~\ref{figridge} when
selecting a
varying number of predictor variables
$p\in\{10^3$, $10^4$, $10^5$, $10^6$, 4,272,227$\}$ as
a consecutive block in the order given by the dataset in
\citet{kogan2009predicting}.
Both the lasso and ridge estimators of the maximin effects are
calculated under the maximal penalty (\ref{eqmax}), which has
computational advantages and avoids having to chose a tuning parameter
for the penalty.

As can be seen from the figures, the variability of
the explained variance is indeed higher for pooled estimation,
compared with the maximin effects estimators, especially for a
lasso-type penalty. The difference in performance between training and
test datasets is also larger for the pooled estimation as it is more
prone to overfitting than the maximin effects estimators.

%

%

\subsection*{Cross-validation} As the dataset is not a priori grouped and
the optimal group size is
unknown,
one needs to decide on the
optimal number of groups $G$. One possibility indicated above is with
cross-validation.
To this end, we split the $n_{\mathrm{train}}=3000$ training data 100 times into
two half-samples of $n_{\mathrm{train}}/2$
observations, sampled uniformly at random. For each split, we then
divide both half-samples
of observations into smaller blocks of consecutive observations. (We
would sample at random if the data did not have a time-ordering.)
The first half-sample of $n_{\mathrm{train}}/2$ observations is split into $G$
blocks with a sample size $n_{\mathrm{train}}/(2G)$
in each block. The second half-sample is split into $g$
blocks, where $g=5$ is chosen as large as possible
while still leaving at least a few hundred observations in each block.
For each split into two half-samples, we compute the maximin estimator on
the blocks formed by the first half-sample and compute the
explained variance in each of the $g$ blocks of the second
half-sample. (The result turns out to be rather insensitive to the precise
choice of $g$; note that we want to choose $g$ as high as possible
but have to keep a minimal number of observations in each of the
``test'' groups in
order not to be overwhelmed by noise in the estimation of the
explained variances.)
The worst-case explained variance over these $g$ groups is then
averaged for each value of $G$ across the 100 random splits into two
half-samples. We choose $G$ to optimize this averaged worst-case explained
variance.
All groups are chosen here as consecutive blocks of equal size from
the two half-samples, respectively, since the reports are ordered
chronologically and
it seems likely that there are shifts in the underlying distributions
over time. If no such time-ordering applies, we would sample the
groups at random within each half-sample for cross-validation.

\subsection*{Computational aspects} 
\begin{figure}[t]

\includegraphics{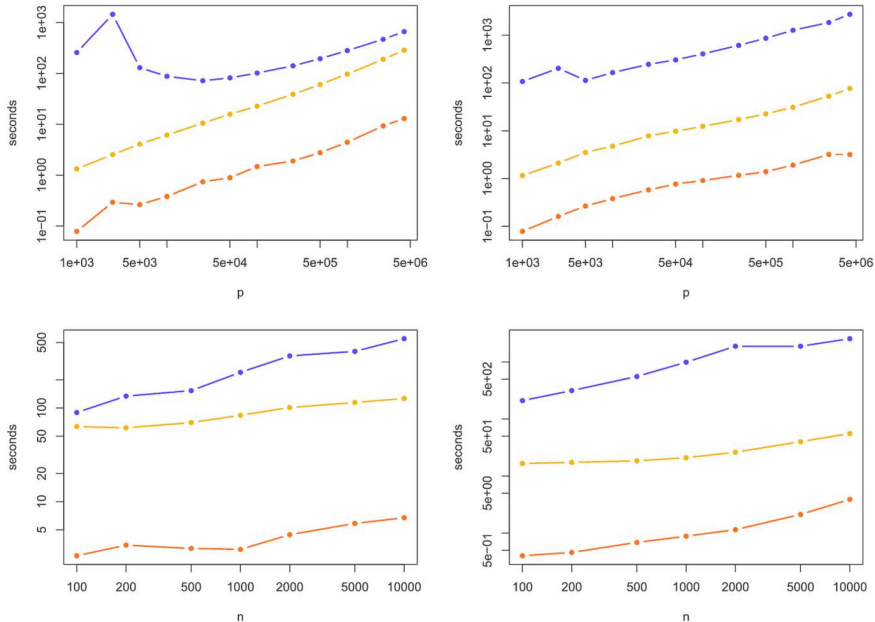}

\caption{The timings in seconds of the three
estimation methods as a function of the number of predictor
variables for $n=3000$ observations for Lasso-type estimation (left)
and ridge
estimation (second from left): the cross-validated pooled estimate
(blue), and
the two estimators of maximin effects with a fixed number of groups
$G=3$ (red) and a cross-validated choice of the number of groups
(orange). Analogous plots for the timings in seconds as a function
of the number of samples for $p=10^6$ variables in the two right
panels.
Estimation of maximin effects was often orders of
magnitudes faster than the pooled estimation.}\label{figtimef}
\end{figure}

Estimation of the standard estimator
(\ref{eqestlambda}) is in
general slower than the pooled estimation over all data, at least as
long as (\ref{eqestlambda}) is computed by iteratively reweighted pooled
estimation. On the other hand, when going for the maximal penalty
estimate as in~(\ref{eqmax}), the solution can be computed using
quadratic or linear programming for ridge and lasso penalties,
respectively, and the design matrix enters only through the inner
products on the right-hand side of (\ref{eqmax}).
Figure~\ref{figtimef} shows the necessary computational time as a
function of the dimensionality $p$ of the data and the number $n$ of
samples.
The advantage of the
maximin effects estimator with a cross-validated choice of the number of
groups is visible across the entire range of the dimensionality. The
relative speed advantage of the maximin estimation is
more than a factor 10 for ridge estimation. Choosing just a fixed
number of groups can get the relative advantage to three orders of
magnitude for ridge estimation, which can be useful in its own right
or as an initial check as to whether there is any signal in the data at all.
The computational complexity is roughly similar as a function of $p$
for the methods whereas the maximin effects have a better scaling as a
function of $n$, as expected since the dimension $n$ drops out of the
memory requirements for estimation and is replaced by the much smaller
number $G$ of groups.

%

\section{Discussion}

One characteristic of large-sale datasets is the mix of a large
number of observations from different sources or different
regimes. Due to the inhomogeneity of the data, estimating
regressions or classifications or graphs over the pool of all
available data is likely to estimate effects that might be strong for
one part of the data but very weak or even of opposite sign for
another part. Here, we proposed to estimate effects which
are present for all possible groupings of the data (even if they
might be masked by noise if we make the groups unreasonably small).
The improvement in predictive accuracy can be seen empirically.

We have introduced the notion of maximin effects and proposed an
estimator for these effects, using either a lasso or ridge-type
penalty. If we have known groups of observations with a different
parameter setting in each group, the estimator is guaranteed to do as
well in estimating the maximin effects as standard Lasso estimation
would in estimating the average effect in a single group of these
observations.
For datasets with unknown groups, we have proposed to sample groups at
random from the available observations.
This has a similar flavor to ``stability selection''
[\citet{meinshausen2008ss}] and the
``bolasso'' [\citet{bach2008bolasso}] or [\citet{bradic2013efficient}],
where models are fitted
repeatedly over random (bootstrap) samples of the data. In contrast to
these approaches, though, the estimator is trying to infer the ``maximin
effects'' if the underlying regression coefficients change randomly,
which is a novel concept. We have presented theoretical guarantees for
the statistical accuracy, an efficient computational algorithm which is
feasible for large-scale problems, as well as empirical results on real
data demonstrating improved performance for prediction.

We expect that the notion
of maximin effects is useful beyond regression and
classification for `big data'' applications, both from a statistical as
well as computational point of view, potentially helping to avoid
detecting too many spurious effects that are not replicable.

\begin{appendix}\label{append}
\section*{Appendix}

\subsection{Proof of Theorem~\texorpdfstring{\protect\ref{theoremmaximin}}{1}}

Let $H$ be its convex hull of the support of $F$.
The maxim effects are given by definition as
\begin{eqnarray*}
b_{\mathrm{maximin}} &=& \mathop{\operatorname{argmax}}_{\beta\in\mathbb{R}^p} \min_{ b
\in F} 2
\beta^t\Sigma b - \beta^t\Sigma\beta
\\
&=& \mathop{\operatorname{argmax}}_{\beta\in\mathbb{R}^p} \min_{ b \in H} 2
\beta^t\Sigma b - \beta^t\Sigma\beta,
\end{eqnarray*}
where the second equality follows by linearity of the objective
function in $b$.

Let $C^tC=\Sigma$ be the Cholesky decomposition of $\Sigma$. Since we
assumed $\Sigma$ to be full-rank, $C$ is invertible, and we define
\[
\tilde{H}:= \{ Cz\dvtx z\in H\} \subseteq\mathbb{R}^p.
\]
Then
%
\begin{equation}
\label{eqhe0} b_{\mathrm{maximin}} = C^{-1} \mathop{\operatorname{argmax}}_{\xi\in\mathbb{R}^p}
\Bigl( \min_{u \in\tilde{H}} 2\xi^t u -\xi^t\xi
\Bigr).
\end{equation}
Define $\xi^*$ to be the choice of weights in the simplex that is
minimizing the $\ell_2$-norm of the corresponding vector in the convex
hull $\tilde{H}$,
%
\begin{equation}
\label{eqgs} \xi^*:= \mathop{\operatorname{argmin}}_{u \in\tilde{H} } \llVert
u\rrVert
_2^2.
\end{equation}
The proof is complete if we can show that
%
\begin{equation}
\label{eq1lts} \xi^* =\mathop{\operatorname{argmax}}_{\xi\in\mathbb{R}^p} \Bigl
( \min
_{u \in\tilde{H}} 2\xi^t u -\xi^t\xi\Bigr),
\end{equation}
since this implies the result, using (\ref{eqhe0}), the invertibility
of $C$ and
\[
C^{-1} \mathop{\operatorname{argmin}}_{u \in\tilde{H} } \llVert u\rrVert
_2^2 = \mathop{\operatorname{argmin}}_{u \in H } u^t
\Sigma u.
\]

By definition of
$\xi^*$ in (\ref{eqgs}), it holds true that for every $\mu\in\tilde
{H} $,
\[
\forall0\le\nu\le1:\qquad\bigl\llVert\xi^* + \nu\bigl( \mu- \xi
^*\bigr)
\bigr\rrVert_2 \ge\bigl\llVert\xi^*\bigr\rrVert_2,
\]
since $\xi^*$ and $\mu$
are from a convex set (namely $\tilde{H}$), and $\xi^*$ minimizes the
$\ell_2$-norm over this convex set.
Since equality holds for $\nu=0$, the derivative of the left-hand side with
respect to $\nu$ at $\nu=0$ has to be positive, which is equivalent to
%
\begin{equation}
\label{eqhe1x} 2 \bigl(\xi^*\bigr)^t \mu- 2 \bigl(\xi^*
\bigr)^t \xi^* \ge0 \qquad\mbox{for all } \mu\in\tilde{H}.
\end{equation}
Hence
%
\begin{equation}
\label{eqhe1} 2 \bigl(\xi^*\bigr)^t \mu- \bigl(\xi^*
\bigr)^t \xi^* \ge\bigl(\xi^*\bigr)^t \xi^* \qquad
\mbox{for all } \mu\in\tilde{H}.
\end{equation}
Choosing $\xi=\xi^*$ yields thus a value of the objective
function of at least $(\xi^*)^t
\xi^*$ in~(\ref{eq1lts}).

On the other hand, choosing $u=\xi^*$ in (\ref{eq1lts}), for all $\xi
\in\mathbb{R}^p$,
%
\begin{equation}
\label{eqhe2} 2\xi^t \xi^* - \xi^t\xi\le\bigl(\xi^*
\bigr)^t\xi^*,
\end{equation}
with equality only if $\xi\equiv\xi^*$. The value of the
objective function in (\ref{eq1lts}) can hence not exceed $ (\xi^*)^t
\xi^* $.
Choosing $\xi=\xi^*$ in (\ref{eq1lts}) yields thus the optimal
value of the objective function and
is indeed a solution to (\ref{eq1lts}), which completes the proof.

\subsection{Proof of Theorem~\texorpdfstring{\protect\ref{theorembasic}}{2}} We write $\hat
{\beta}$
instead of
$\hat{\beta}^\kappa$ to simplify notation, and
use the constrained
version of the estimator. Note that the theorem is for known
groups, and thus $\{b\dvtx b\in F\}= \{b\dvtx b=b_g$ for some
$g\in\{1.\ldots,G\}\}$.
Let $\delta_g = (G/n) X_g^t \varepsilon_g$ for all $g\in\{1,\ldots,G\}$.
Using the basic inequality, for any fixed vector $\xi$
in the feasible region, that is, for all $\xi$ with $\llVert \xi
\rrVert _1\le\kappa$,
%
\begin{eqnarray}
\qquad \min_g \bigl\{2\hat{\beta}^t \hat{
\Sigma}_g b_g - \hat{\beta}^t\hat{
\Sigma}_g \hat{\beta} + 2\hat{\beta}^t
\delta_g \bigr\} &\ge& \min_g \bigl\{ 2
\xi^t \hat{\Sigma}_g b_g - \xi^t
\hat{\Sigma}_g \xi+2 \xi^t \delta_g \bigr\}
\label{eqbasic}.
\end{eqnarray}
Using the definition of $D=\max_g\llVert \hat{\Sigma}_g-\Sigma\rrVert
_\infty$,
%
\begin{eqnarray}
\qquad && \min_g \bigl\{2\hat{\beta}^t \Sigma
b_g - \hat{\beta}^t\Sigma\hat{\beta} \bigr\} +2D\llVert
\hat{\beta}\rrVert_1 \max_g\llVert
b_g\rrVert_1 + D\llVert\hat{\beta}\rrVert
_1^2 + 2\llVert\hat{\beta}\rrVert_1
\llVert\delta_g\rrVert_\infty\label{eqbasic2aa}
\\
&&\qquad \ge \min_g \bigl\{ 2 \xi^t \Sigma b_g
- \xi^t\Sigma\xi\bigr\}
\nonumber\\[-8pt]\label{eqbasic2} \\[-8pt]\nonumber
&&\quad\qquad{}  -2D\llVert\xi\rrVert_1 \max
_g\llVert b_g\rrVert_1 - D\llVert
\xi\rrVert_1^2 -2 \llVert\xi\rrVert_1
\llVert\delta_g\rrVert_\infty.\nonumber
\end{eqnarray}
Hence, using $\xi=b_{\mathrm{maximin}}$, and using that by definition of
$\hat{\beta}$, $\kappa\ge\break \max\{\max_g\llVert b_g\rrVert _1,
\llVert \hat{\beta}\rrVert _1
\}$ (and hence, when using Theorem~\ref{theoremmaximin} that
$b_{\mathrm{maximin}}$ is in the convex hull of $F_B$, also $\kappa\ge\llVert
b_{\mathrm{maximin}}\rrVert _1$), it follows that
%
\begin{equation}\label{eqbasic3}
\min_g \bigl\{2\hat{\beta}^t \Sigma
b_g - \hat{\beta}^t\Sigma\hat{\beta} \bigr\} \ge V^*-6D
\kappa^2 -4 \max_g \llVert\delta_g
\rrVert_\infty\kappa,
\end{equation}
where
\[
V^* = \min_g \bigl\{ 2 b_{\mathrm{maximin}}^t
\Sigma b_g - b_{\mathrm{maximin}}^t\Sigma b_{\mathrm{maximin}}
\bigr\},
\]
which completes the proof.

\subsection{Proof of Theorem~\texorpdfstring{\protect\ref{theoremrandom}}{3}}
Starting as in the proof of Theorem~\ref{theorembasic}, let
$I_g\subseteq\{1,\ldots,n\}$ be the index set of
the $g$th group. The explained variance in group $g$ when using a
regression vector $\xi\in\mathbb{R}^p$ can then be
written as
\[
\sum_{i\in I_g} 2(X_i \xi) (X_i
B_i) - \sum_{i\in I_g} (X_i
\xi)^2.
\]
Analogous to (\ref{eqbasic}), we have the basic inequality for all
$\xi$ with $\llVert \xi\rrVert _1\le\kappa$,
%
\begin{eqnarray}\label{eqbasic2a}
&& \min_g \frac{1} m \sum
_{i\in I_g} \bigl( 2(X_i \hat{\beta}) (X_i
B_i) - (X_i\hat{\beta})^2 +
2(X_i \hat{\beta}) \varepsilon_i \bigr)
\nonumber\\[-8pt]\\[-8pt]\nonumber
&&\qquad \ge\min
_g \frac{1} m \sum_{i\in
I_g}\bigl( 2(X_i \xi) (X_i B_i) -
(X_i\xi)^2 + 2(X_i \xi)
\varepsilon_i \bigr).
\end{eqnarray}
As the Pareto condition is fulfilled (with corresponding probability
$1-\gamma$), there exists a subset
$\tilde{G}\subseteq\{1,\ldots,G\}$ such that $B_i=b_g$ for all $i\in
I_g, g\in\tilde{G}$, and an essential subset is formed by
$A=\{b_g;g\in\tilde{G}\}$. Restricting the minimum on the left-hand
side of
(\ref{eqbasic2a}) over all groups in $\tilde{G}$, we have for all
$\xi$ with $\llVert \xi\rrVert _1\le\kappa$,
\begin{eqnarray*}
&& \min_{g\in\tilde{G}} \bigl(2\hat{\beta}^t \hat{
\Sigma}_g b_g - \hat{\beta}^t\hat{
\Sigma}_g\hat{\beta} +2\hat{\beta}^t \delta_g
\bigr)
\\
&&\qquad \ge \min_g \frac{1} m \sum
_{i\in I_g} \bigl( 2(X_i \xi) (X_i
B_i) - (X_i\xi)^2 + 2(X_i \xi)
\varepsilon_i \bigr),
\end{eqnarray*}
where $\delta_g= (1/m) X_g^t \varepsilon_g$. Hence, using $\llVert \xi
\rrVert _1\le\kappa$,
\begin{eqnarray*}
\min_{g\in\tilde{G}} \bigl\{2\hat{\beta}^t \hat{
\Sigma}_g b_g - \hat{\beta}^t\hat{
\Sigma}_g\hat{\beta} \bigr\} +4 \kappa\llVert\delta_g
\rrVert_\infty&\ge&\min_g \frac{1} m \sum
_{i\in
I_g} \bigl( 2(X_i \xi)
(X_i B_i) - (X_i\xi)^2 \bigr).
\end{eqnarray*}
Analogous to (\ref{eqbasic2aa}), the first term on the left-hand side
is bounded with probability at least $1-\alpha$ by
\[
\min_{g\in\tilde{G}} \bigl\{2\hat{\beta}^t \hat{
\Sigma}_g b_g - \hat{\beta}^t\hat{
\Sigma}_g\hat{\beta} \bigr\} \le\min_{g\in\tilde{G}} \bigl\{2
\hat{\beta}^t \Sigma b_g - \hat{\beta}^t\Sigma
\hat{\beta} \bigr\} + 3 D \kappa^2.
\]
Since $A$ is by assumption an essential subset, we have that the first
term on the left-hand side is bounded with probability at least
$1-\alpha$ by
\[
\min_{g\in\tilde{G}} \bigl\{2\hat{\beta}^t \hat{
\Sigma}_g b_g - \hat{\beta}^t\hat{
\Sigma}_g\hat{\beta} \bigr\} \le\min_{b\in F}
V_{\hat{\beta};b} + 3 D \kappa^2.
\]
Thus, for all $\xi$ with $\llVert \xi\rrVert _1\le\kappa$,
\begin{eqnarray*}
\min_{b\in F} V_{\hat{\beta};b}+4 \kappa\llVert
\delta_g\rrVert_\infty+ 3 D \kappa^2 &\ge&\min
_g \frac{1} m \sum_{i\in
I_g}
\bigl( 2(X_i \xi) (X_i B_i) -
(X_i\xi)^2 \bigr).
\end{eqnarray*}
Since $b_{\mathrm{maximin}}$ is in the feasible region, we can use it for $\xi$
to get
%
\begin{eqnarray}\label{eqaddh1}
&& \min_{b\in F} V_{\hat{\beta};b}+4 \kappa\llVert
\delta_g\rrVert_\infty+ 3 D \kappa^2
\nonumber\\[-8pt]\\[-8pt]\nonumber
&&\qquad \ge \min_g \frac{1} m \sum_{i\in
I_g}\bigl( 2(X_i b_{\mathrm{maximin}}) (X_i B_i) -
(X_i b_{\mathrm{maximin}})^2 \bigr).
\end{eqnarray}

Now, by definition of $b_{\mathrm{maximin}}$, when letting $H$ be the convex
hull of $F$ (where $F$ is again the support of $F_B$),
\begin{eqnarray*}
&& \min_{b\in F} E \bigl( 2 (X_i b_{\mathrm{maximin}})
(X_i b) - (X_i b_{\mathrm{maximin}})^2 \bigr)
\\
&&\qquad = \min_{b\in H}E \bigl( 2 (X_i b_{\mathrm{maximin}})
(X_i b) - (X_i b_{\mathrm{maximin}})^2 \bigr)
\\
&&\qquad = E \bigl( (X_i b_{\mathrm{maximin}})^2 \bigr) =
b_{\mathrm{maximin}}^t \Sigma b_{\mathrm{maximin}} = V^*,
\end{eqnarray*}
where we have used in the first equality linearity with respect to the
argument $b\in F$ and in the second the definition of $b_{\mathrm{maximin}}$
and the fact that $b_{\mathrm{maximin}}$ is in the convex hull of the support
$F$ of
$F_B$.

Now bounding the fluctuations on the right-hand side of (\ref
{eqaddh1}), we use that $|(X_i \xi) (X_i b -X_i\xi)|\le
\llVert X\rrVert _\infty^2 (\max_{b\in F}\llVert b\rrVert _1
+\llVert \xi\rrVert _1) \llVert \xi\rrVert _1 \le2 \kappa^2$.
Using Hoeffding's inequality and a union bound over all groups, for
any $\alpha\in(0,1)$, if $\kappa\ge\max_g\llVert b_g\rrVert _1$ and
$\kappa\ge\llVert \xi\rrVert _1$,
\begin{eqnarray*}
\hspace*{-5.5pt}&& P \biggl( \min_g \frac{1}{m} \sum
_{i\in I_g} \bigl( 2(X_i \xi) (X_i
B_i) - (X_i\xi)^2 \bigr) \ge
b_{\mathrm{maximin}}^t \Sigma b_{\mathrm{maximin}} - \frac{\sqrt{2\log(G/\alpha)}}{\sqrt{m}}
\kappa^2 \biggr)
\\
\hspace*{-5.5pt}&&\qquad  \ge1- \alpha.
\end{eqnarray*}
Plugging this into (\ref{eqaddh1}), it holds with probability
$1-2\alpha$ for all $\xi$ with $\llVert \xi\rrVert _1\le\kappa$ that
%
\begin{eqnarray}
\label{eqbasic3c} \min_{b\in F} V_{\hat{\beta};b} &\ge& V^* -
\frac{\sqrt{2\log(G/\alpha)}}{\sqrt{m}}\kappa^2 -4 \kappa\llVert\delta_g
\rrVert_\infty- 3 D \kappa^2,
\end{eqnarray}
which shows the first part of the claim in (\ref{eqtheorand}) if we use a
union bound to exclude the event which does not correspond to the
Pareto condition, and hence this excluded event has corresponding
probability at most $\gamma$. The value of $D$ can be bounded with the
help of (\ref{eqdeviation}) with probability $1-\alpha$ to yield
\[
\min_{b\in F} V_{\hat{\beta};b} \ge V^* - 6 \frac{\sqrt{2\log(2 p^2
G/\alpha)}}{\sqrt{m}}
\kappa^2 -4 \llVert\delta_g\rrVert_\infty
\kappa,
\]
and the latter bound will then hold true with probability at least
$1-3\alpha-\gamma$.
The second part in (\ref{eqtheorand}) follows as (\ref{eqbasic3c}) implies
%
\begin{equation}
2\hat{\beta}^t \Sigma b_{\mathrm{maximin}} - \hat{\beta}^t
\Sigma\hat{\beta} \ge V^* - 6 \frac{\sqrt{2\log(2 p^2 G/\alpha)}}{\sqrt
{m}}\kappa^2 -4 \llVert
\delta_g\rrVert_\infty\kappa.
\end{equation}
The claim follows by $V^*=b_{\mathrm{maximin}}^t \Sigma b_{\mathrm{maximin}} $ and
rearranging terms.
\end{appendix}



%

\printaddresses
\end{document}